\documentclass[final,3p,authoryear,onecolumn]{elsarticle}  

\usepackage{amssymb}
\usepackage{xcolor}
\usepackage{tcolorbox}
\usepackage[framemethod=TikZ]{mdframed}
\usepackage{lipsum}
\usepackage{pdfpages}
\usepackage{pdflscape}
\mdfdefinestyle{MyFrame}{%
	linecolor=blue,
	outerlinewidth=2pt,
	roundcorner=20pt,
	innertopmargin=\baselineskip,
	innerbottommargin=\baselineskip,
	innerrightmargin=10pt,
	innerleftmargin=10pt,
	backgroundcolor=gray!50!white}

\journal{ESO}

\begin{document}

\begin{frontmatter}

\title{Sailing to the next safe harbour in our trip to the early Universe:\\ The massive star population of metal-poor galaxies }

\author{N. Castro$^1$, M. Garcia$^2$, A. Herrero$^{3,4}$,  A. A. C. Sander$^{5}$, A. F. McLeod$^{6,7}$, M. M. Roth$^1$,  I. Negueruela$^{8,9}$ and J. S. Vink$^{10}$}

\address{$^1$Leibniz-Institut f\"ur Astrophysik Potsdam, An der Sternwarte 16, D-14482 Potsdam, Germany}
\address{$^2$Centro de Astrobiolog\'ia, CSIC-INTA. Crtra. de Torrej\'on a Ajalvir km 4., E-28850 Torrej\'on de Ardoz (Madrid), Spain}
\address{$^3$Departamento de Astrofísica, Universidad de La Laguna, 38205, La Laguna, Tenerife, Spain}
\address{$^4$Instituto de Astrofísica de Canarias, 38200, La Laguna, Tenerife, Spain}
\address{$^5$Zentrum für Astronomie der Universität Heidelberg, Astronomisches Rechen-Institut, Mönchhofstr. 12–14, 69120, Heidelberg, Germany}
\address{$^6$Centre for Extragalactic Astronomy, Department of Physics, Durham University, Durham, UK}
\address{$^7$Institute for Computational Cosmology, Department of Physics, University of Durham, Durham, UK}
\address{$^8$Departamento de Física Aplicada, Facultad de Ciencias, Universidad de Alicante, Carretera de San Vicente s/n, 03690 San Vicente del Raspeig, Spain}
\address{$^9$Instituto Universitario de Investigación Informática, Universidad de Alicante, San Vicente del Raspeig, Spain}
\address{$^{10}$Armagh Observatory, College Hill, Armagh, BT61 9DG Northern Ireland, UK}

\begin{abstract}

Very metal-poor massive stars in the Local Group are our best proxies for the Universe’s first stars, making them essential for modeling reionization and early galactic chemical evolution. Studying such stars in our Local Universe is key to extrapolating our knowledge to more distant regions, where individual massive stars cannot be resolved but are dynamically and chemically shaping their environments. The MUSE integral field spectrograph has transformed massive star studies in the Milky Way and Magellanic Clouds, but resolving star-forming galaxies containing very metal-poor stars is at the limit of the current field of view and sensitivity. Therefore, only small dedicated efforts of selected regions are studied, providing us with snapshots of low-metallicity massive stars rather than a comprehensive picture. This scarcity is a major bottleneck for understanding and sufficiently modelling the evolution and feedback of massive stars across cosmic time. We therefore envision a new generation of panoramic integral-field spectrographs and high multiplex multi-object spectrographs mounted on dedicated large optical telescopes. Such facilities will not only allow to resolve very-metal-pool galaxies, but further enable the systematic exploration of the massive stellar content across the entire Local Group, thereby reaching a new era in massive star studies and understanding\footnote{The abstract has been added to the arXiv version.}.

\end{abstract}

	


\end{frontmatter}

\newpage
\section{Probing massive star evolution at metallicities below the Magellanic Clouds}

Galaxies are chemically and dynamically shaped by massive stars (M$>$8\,M$\odot$). As principal sources of heavy elements and UV radiation, massive stars play a fundamental role in the composition and ionization of the Universe \citep{2012ARA&A..50..107L}. With their final core collapse, massive stars are the most plausible progenitors of gamma-ray bursts and various types of supernovae \citep[e.g][]{2006ARA&A..44..507W}.  The strong mass loss during their life as well as their final fate changes the surrounding interstellar medium (ISM) and defines the spectrum of the gravitational wave events we measure today \citep{2025arXiv250818082T}. Understanding star-forming galaxies in our local neighborhood, interpreting the observations at the peak of star formation in the Universe  \citep[z$\sim$2,][and references therein]{2014ARA&A..52..415M} and explaining the reionization epoch require to comprehend the formation and evolution of massive stars under very different physical conditions. However, the formation and evolution of massive stars are far from being understood. Stellar evolution is mainly controlled by the initial mass of the stars, but other factors shape and change their evolutionary paths. Metallicity, rotational velocity, strong stellar winds, binary interaction, magnetic fields and mergers affect the evolutionary channels and the lifetime of a star \citep{2000ARA&A..38..143M,2012ARA&A..50..107L,2000ARA&A..38..613K,2016MNRAS.457.2355S}. The role of these parameters is both more important and uncertain for the most massive stars \citep[$>$100\,M$\odot$,][]{2015HiA....16...51V}.

\medskip

\begin{figure}[!h] 
	\resizebox{\hsize}{!}{\includegraphics[angle=0,width=0.1\textwidth]{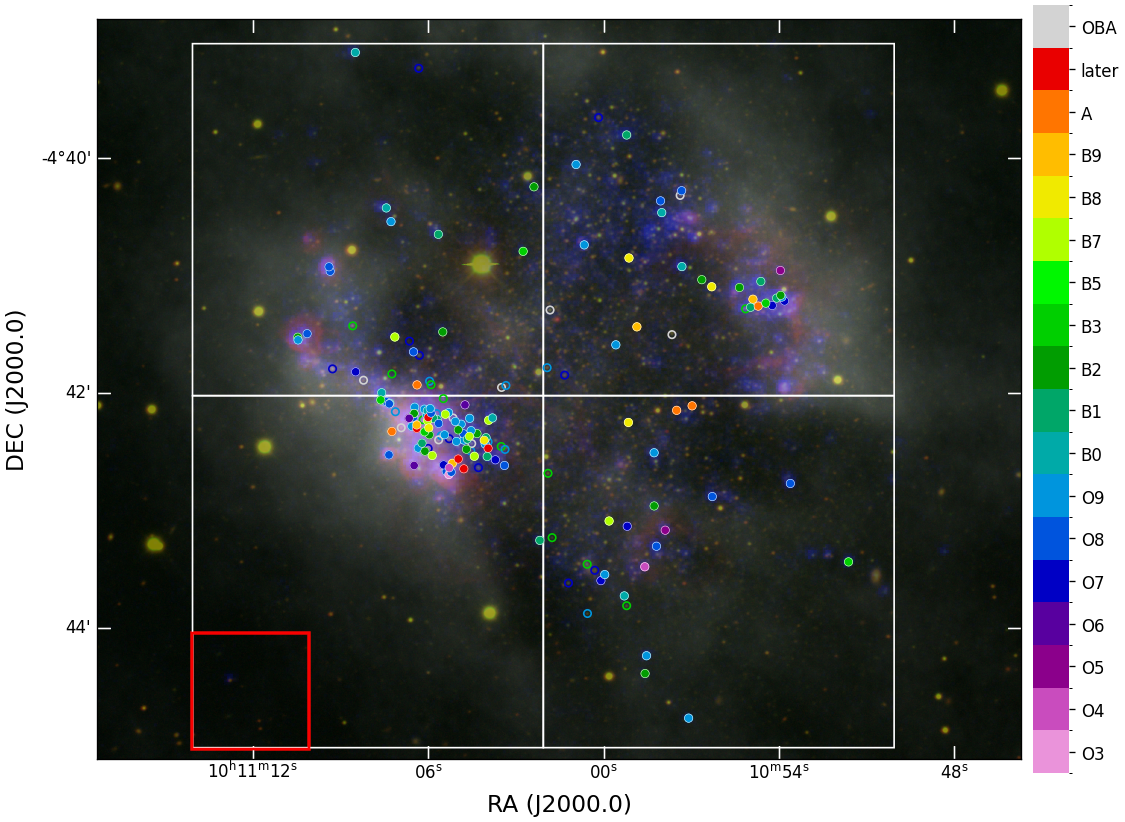}}
	\caption{   The 1/10\,Z$\odot$ dwarf galaxy Sextans A, along with the OB-type stars identified by \cite{2022MNRAS.516.4164L}, is shown. The field of view of MUSE (red square) is compared with that of a potential next-generation integral-field spectrograph featuring a large panoramic 3'$\times$3' field of view (white squares). With such a facility, only four pointings would be required to complete a census of the galaxy’s massive star population and to derive stellar parameters for all of them, thereby establishing a new standard for low-Z massive star evolution in an exceptionally modest amount of observing time.}
	\label{Fig:sextA} \end{figure}

In a Universe where the average chemical composition increases over time, understanding how massive stars behave as a function of metallicity (Z), and particularly in metal-poor (low-Z) environments, is essential. The Small Magellanic Cloud (SMC; 1/5\,Z$\odot$) has traditionally served as the benchmark for low-Z studies. However, its metallicity is representative only of the relatively late Universe. Consequently, the insights gained from SMC massive stars do not capture the conditions that prevailed during key cosmic epochs, such as the peak of the cosmic star-formation rate or earlier periods. To probe these regimes, we must reach to even lower metallicities. Although theoretical models can simulate stars at such low-Z, the number of unconstrained parameters increases as metallicity decreases, introducing significant uncertainties into the extrapolations used to model unresolved stellar populations \citep{2015A&A...581A..15S,2025A&A...703A.131S}.

\medskip

We need large systematic surveys to unveil the evolution and nature of massive stars and the role of metallicity. Systematic studies of complete populations will provide homogeneous results and unbiased empirical anchors for the theory of stellar evolution as they have done, for instance, for the Milky Way and 30\,Doradus \citep{2011A&A...530A.108E,2014A&A...570L..13C,2022A&A...665A.150H}. These homogeneous analyses of large samples shed new light on mixing mechanisms at stellar interiors, unveiled binary interaction products, and enabled quantifying the dependence of wind mass loss on metallicity  \citep{2007A&A...473..603M,2008ApJ...676L..29H,2025A&A...693L..10M}, but only down to the 1/5\,Z$\odot$ SMC metallicity.

\medskip

Within the Local Group, several galaxies are known to have metal contents lower than those of the Magellanic Clouds and to host star-formation bursts with detected massive stars \citep{2021ExA....51..887G}. Sextans A is a clear example. \citet{2022MNRAS.516.4164L} unveiled its population of OB stars in this dwarf galaxy with a metallicity of only 1/10\,Z$\odot$, making it an excellent environment in which to pursue studies of massive star evolution at very low metallicity (Fig.~\ref{Fig:sextA}). Other promising galaxies, such as SagDIG (1/20\,Z$\odot$), Leo P (1/30\,Z$\odot$), and the iconic blue-compact dwarf I Zw 18 (1/32\,Z$\odot$), are also under close scrutiny. However, providing a systematic and homogeneous census of massive stars in these systems, and building archaeological evolutionary maps \citep{2014A&A...570L..13C}, remains observationally challenging with current instrumentation. It becomes nearly impossible when aiming to characterize individual stars at the distance of I Zw 18 (18.9\,Mpc; \citealt{1998ApJ...508..248V}), although indirect approaches based on unresolved stellar populations can still be explored \citep[e.g.,][]{2017ApJ...845..165M}.

\medskip

\section{The next generations of integral field spectrographs}

State-of-the-art integral field spectrographs have shown the maturity of the technology and the potential to deblend crowded stellar fields \citep{2013A&A...549A..71K,2018A&A...618A...3R}, holding great promise to cover full populations of massive stars beyond the Magellanic Clouds. MUSE \citep{2014Msngr.157...13B} has provided efficient ways to systematically study stellar dense clusters impossible to reach, or too expensive, with other facilities \citep[e.g.,][]{2021A&A...648A..65C}.  However, the field-of-view of MUSE is still not ideal to map large star-forming associations. The next generation of multi-object spectrographs will be crucial for probing the stellar content beyond the dense cores of clusters, but they are still not sufficient to carry out a homogeneous spectroscopic study of large massive stellar population. For example, WEAVE and 4MOST \citep{2024MNRAS.530.2688J,2019Msngr.175....3D} will be fundamental to build a complete picture of the stellar evolution of the most massive stars in the Milky Way and Magellanic Clouds. However,  pushing this analysis to lower metallicities requires the use of $\gtrsim$\,8-m–class telescopes \citep[e.g.,][]{2021ExA....51..887G}, as galaxies with metallicities poorer than that of the SMC are located at distances of at least $\sim$750 kpc.

\medskip

A new generation of large multiplex multi-object and wide-field panoramic integral field spectrographs  on state-of-the-art telescopes is essential for conducting extensive, systematic spectroscopic surveys (e.g., Fig.~\ref{Fig:sextA}). Such a facility will enable the first comprehensive study of massive stellar populations at metallicities comparable to or below that of the SMC, with the following key goals:

\begin{itemize}
\item Quantify the influence of metallicity on the evolution of massive stars and establish its connection to the formation and evolution of the earliest stellar populations.
\item Identify and characterize spectroscopic binaries to investigate the role of multiplicity in shaping massive star evolution across environments with varying metallicity.
\item Search for empirical evidences of chemically homogeneous evolution in massive stars at low metallicity.
\item Investigate the role of metallicity in the upper stellar mass limit.
\item Use continued spectroscopic mapping to characterize the interaction between massive stars and the surrounding interstellar medium in active star-forming regions \citep{2020ApJ...891...25M}. These low-metallicity Local Group galaxies are low-mass systems in which stellar feedback strongly regulates star formation and governs ionizing-photon escape fractions, providing a direct link to conditions in the early universe.

\end{itemize}

Dedicated large telescopes optimized for spectroscopic surveys would provide unique capabilities for stellar astrophysics and enable strong synergies with current facilities, such as the Vera Rubin Observatory, as well as with future ground- and space-based missions (e.g. the Nancy Grace Roman Space Telescope). This represents a transformative concept that will allow us to study the massive stellar populations of galaxies at the outer reaches of the Local Group and beyond in low-metallicity environments.

\bibliographystyle{aa}
\setlength{\bibsep}{0pt}
\bibliography{WP}

@Article{2012ARA&A..50..107L,
  author        = {{Langer}, N.},
  title         = {{Presupernova Evolution of Massive Single and Binary Stars}},
  journal       = {Annual Review of Astronomy and Astrophysics},
  year          = {2012},
  volume        = {50},
  pages         = {107-164},
  month         = Sep,
  adsurl        = {https://ui.adsabs.harvard.edu/#abs/2012ARA&A..50..107L},
  archiveprefix = {arXiv},
  doi           = {10.1146/annurev-astro-081811-125534},
  eprint        = {1206.5443},
  primaryclass  = {astro-ph.SR},
}

@Article{2014A&A...570L..13C,
  author        = {{Castro}, N. and {Fossati}, L. and {Langer}, N. and {Sim{\'o}n-D{\'\i}az}, S. and {Schneider}, F.~R.~N. and {Izzard}, R.~G.},
  title         = {{The spectroscopic Hertzsprung-Russell diagram of Galactic massive stars}},
  journal       = {A\&A},
  year          = {2014},
  volume        = {570},
  pages         = {L13},
  month         = Oct,
  adsurl        = {https://ui.adsabs.harvard.edu/#abs/2014A&A...570L..13C},
  archiveprefix = {arXiv},
  doi           = {10.1051/0004-6361/201425028},
  eid           = {L13},
  eprint        = {1410.3499},
  primaryclass  = {astro-ph.SR},
}

@Article{2006ARA&A..44..507W,
  author        = {{Woosley}, S.~E. and {Bloom}, J.~S.},
  title         = {{The Supernova Gamma-Ray Burst Connection}},
  journal       = {Annu. Rev. Astron. Astrophys.},
  year          = {2006},
  volume        = {44},
  number        = {1},
  pages         = {507-556},
  month         = {Sep},
  adsurl        = {https://ui.adsabs.harvard.edu/abs/2006ARA&A..44..507W},
  archiveprefix = {arXiv},
  doi           = {10.1146/annurev.astro.43.072103.150558},
  eprint        = {astro-ph/0609142},
  primaryclass  = {astro-ph},
}

@Article{2011A&A...530A.108E,
  author        = {{Evans}, C.~J. and {Taylor}, W.~D. and {H{\'e}nault-Brunet}, V. and {Sana}, H. and {de Koter}, A. and {Sim{\'o}n-D{\'{\i}}az}, S. and {Carraro}, G. and {Bagnoli}, T. and {Bastian}, N. and {Bestenlehner}, J.~M. and {Bonanos}, A.~Z. and {Bressert}, E. and {Brott}, I. and {Campbell}, M.~A. and {Cantiello}, M. and {Clark}, J.~S. and {Costa}, E. and {Crowther}, P.~A. and {de Mink}, S.~E. and {Doran}, E. and {Dufton}, P.~L. and {Dunstall}, P.~R. and {Friedrich}, K. and {Garcia}, M. and {Gieles}, M. and {Gr{\"a}fener}, G. and {Herrero}, A. and {Howarth}, I.~D. and {Izzard}, R.~G. and {Langer}, N. and {Lennon}, D.~J. and {Ma{\'{\i}}z Apell{\'a}niz}, J. and {Markova}, N. and {Najarro}, F. and {Puls}, J. and {Ramirez}, O.~H. and {Sab{\'{\i}}n-Sanjuli{\'a}n}, C. and {Smartt}, S.~J. and {Stroud}, V.~E. and {van Loon}, J.~T. and {Vink}, J.~S. and {Walborn}, N.~R.},
  title         = {{The VLT-FLAMES Tarantula Survey. I. Introduction and observational overview}},
  journal       = {A\&A},
  year          = {2011},
  volume        = {530},
  pages         = {A108},
  month         = jun,
  adsurl        = {http://adsabs.harvard.edu/abs/2011A%26A...530A.108E},
  archiveprefix = {arXiv},
  doi           = {10.1051/0004-6361/201116782},
  eid           = {A108},
  eprint        = {1103.5386},
  primaryclass  = {astro-ph.SR},
}

@Article{2014Msngr.157...13B,
  author  = {{Bacon}, R. and {Vernet}, J. and {Borisova}, E. and {Bouch{\'e}}, N. and {Brinchmann}, J. and {Carollo}, M. and {Carton}, D. and {Caruana}, J. and {Cerda}, S. and {Contini}, T. and {Franx}, M. and {Girard}, M. and {Guerou}, A. and {Haddad}, N. and {Hau}, G. and {Herenz}, C. and {Herrera}, J.~C. and {Husemann}, B. and {Husser}, T.-O. and {Jarno}, A. and {Kamann}, S. and {Krajnovic}, D. and {Lilly}, S. and {Mainieri}, V. and {Martinsson}, T. and {Palsa}, R. and {Patricio}, V. and {P{\'e}contal}, A. and {Pello}, R. and {Piqueras}, L. and {Richard}, J. and {Sandin}, C. and {Schroetter}, I. and {Selman}, F. and {Shirazi}, M. and {Smette}, A. and {Soto}, K. and {Streicher}, O. and {Urrutia}, T. and {Weilbacher}, P. and {Wisotzki}, L. and {Zins}, G.},
  title   = {{MUSE Commissioning}},
  journal = {The Messenger},
  year    = {2014},
  volume  = {157},
  pages   = {13-16},
  month   = sep,
  adsurl  = {http://adsabs.harvard.edu/abs/2014Msngr.157...13B},
}

@Article{2016MNRAS.457.2355S,
  author        = {{Schneider}, F.~R.~N. and {Podsiadlowski}, P. and {Langer}, N. and {Castro}, N. and {Fossati}, L.},
  title         = {{Rejuvenation of stellar mergers and the origin of magnetic fields in massive stars}},
  journal       = {MNRAS},
  year          = {2016},
  volume        = {457},
  pages         = {2355-2365},
  month         = apr,
  adsurl        = {http://adsabs.harvard.edu/abs/2016MNRAS.457.2355S},
  archiveprefix = {arXiv},
  doi           = {10.1093/mnras/stw148},
  eprint        = {1601.05084},
  primaryclass  = {astro-ph.SR},
}

@Article{2000ARA&A..38..143M,
  author        = {{Maeder}, Andr{\'e} and {Meynet}, Georges},
  title         = {{The Evolution of Rotating Stars}},
  journal       = {ARA\&A},
  year          = {2000},
  volume        = {38},
  pages         = {143-190},
  month         = Jan,
  adsurl        = {https://ui.adsabs.harvard.edu/#abs/2000ARA&A..38..143M},
  archiveprefix = {arXiv},
  doi           = {10.1146/annurev.astro.38.1.143},
  eprint        = {astro-ph/0004204},
  primaryclass  = {astro-ph},
}

@Article{2015HiA....16...51V,
  author        = {{Vink}, Jorick S. and {Heger}, Alexander and {Krumholz}, Mark R. and {Puls}, Joachim and {Banerjee}, S. and {Castro}, N. and {Chen}, K. -J. and {Chen{\`e}}, A. -N. and {Crowther}, P.~A. and {Daminelli}, A. and {Gr{\"a}fener}, G. and {Groh}, J.~H. and {Hamann}, W. -R. and {Heap}, S. and {Herrero}, A. and {Kaper}, L. and {Najarro}, F. and {Oskinova}, L.~M. and {Roman-Lopes}, A. and {Rosen}, A. and {Sander}, A. and {Shirazi}, M. and {Sugawara}, Y. and {Tramper}, F. and {Vanbeveren}, D. and {Voss}, R. and {Wofford}, A. and {Zhang}, Y.},
  title         = {{Very Massive Stars in the local Universe}},
  journal       = {Highlights of Astronomy},
  year          = {2015},
  volume        = {16},
  pages         = {51-79},
  month         = Mar,
  adsurl        = {https://ui.adsabs.harvard.edu/#abs/2015HiA....16...51V},
  archiveprefix = {arXiv},
  doi           = {10.1017/S1743921314004657},
  eprint        = {1302.2021},
  primaryclass  = {astro-ph.SR},
}

@Article{2014ARA&A..52..415M,
  author        = {{Madau}, Piero and {Dickinson}, Mark},
  title         = {{Cosmic Star-Formation History}},
  journal       = {ARA\&A},
  year          = {2014},
  volume        = {52},
  pages         = {415-486},
  month         = {Aug},
  adsurl        = {https://ui.adsabs.harvard.edu/abs/2014ARA&A..52..415M},
  archiveprefix = {arXiv},
  doi           = {10.1146/annurev-astro-081811-125615},
  eprint        = {1403.0007},
  primaryclass  = {astro-ph.CO},
}

@Article{2013A&A...549A..71K,
  author        = {{Kamann}, S. and {Wisotzki}, L. and {Roth}, M.~M.},
  title         = {{Resolving stellar populations with crowded field 3D spectroscopy}},
  journal       = {A\&A},
  year          = {2013},
  volume        = {549},
  pages         = {A71},
  month         = {Jan},
  adsurl        = {https://ui.adsabs.harvard.edu/abs/2013A&A...549A..71K},
  archiveprefix = {arXiv},
  doi           = {10.1051/0004-6361/201220476},
  eid           = {A71},
  eprint        = {1211.0445},
  primaryclass  = {astro-ph.IM},
}

@Article{2018A&A...618A...3R,
  author        = {{Roth}, Martin M. and {Sandin}, Christer and {Kamann}, Sebastian and {Husser}, Tim-Oliver and {Weilbacher}, Peter M. and {Monreal-Ibero}, Ana and {Bacon}, Roland and {den Brok}, Mark and {Dreizler}, Stefan and {Kelz}, Andreas and {Marino}, Raffaella Anna and {Steinmetz}, Matthias},
  title         = {{MUSE crowded field 3D spectroscopy in NGC 300. I. First results from central fields}},
  journal       = {A\&A},
  year          = {2018},
  volume        = {618},
  pages         = {A3},
  month         = {Oct},
  adsurl        = {https://ui.adsabs.harvard.edu/abs/2018A&A...618A...3R},
  archiveprefix = {arXiv},
  doi           = {10.1051/0004-6361/201833007},
  eid           = {A3},
  eprint        = {1806.04280},
  primaryclass  = {astro-ph.GA},
}

@ARTICLE{2022A&A...665A.150H,
       author = {{Holgado}, G. and {Sim{\'o}n-D{\'\i}az}, S. and {Herrero}, A. and {Barb{\'a}}, R.~H.},
        title = "{The IACOB project. VII. The rotational properties of Galactic massive O-type stars revisited}",
      journal = {A\&A},
         year = 2022,
        month = sep,
       volume = {665},
          eid = {A150},
        pages = {A150},
          doi = {10.1051/0004-6361/202243851},
archivePrefix = {arXiv},
       eprint = {2207.12776},
 primaryClass = {astro-ph.SR},
       adsurl = {https://ui.adsabs.harvard.edu/abs/2022A&A...665A.150H}
}

@ARTICLE{2000ARA&A..38..613K,
       author = {{Kudritzki}, Rolf-Peter and {Puls}, Joachim},
        title = "{Winds from Hot Stars}",
      journal = {Annual Review of Astronomy and Astrophysics},
         year = 2000,
        month = jan,
       volume = {38},
        pages = {613-666},
          doi = {10.1146/annurev.astro.38.1.613},
       adsurl = {https://ui.adsabs.harvard.edu/abs/2000ARA&A..38..613K}
}

@ARTICLE{2008ApJ...676L..29H,
       author = {{Hunter}, I. and {Brott}, I. and {Lennon}, D.~J. and {Langer}, N. and {Dufton}, P.~L. and {Trundle}, C. and {Smartt}, S.~J. and {de Koter}, A. and {Evans}, C.~J. and {Ryans}, R.~S.~I.},
        title = "{The VLT FLAMES Survey of Massive Stars: Rotation and Nitrogen Enrichment as the Key to Understanding Massive Star Evolution}",
      journal = {ApJL},
         year = 2008,
        month = mar,
       volume = {676},
       number = {1},
        pages = {L29},
          doi = {10.1086/587436},
archivePrefix = {arXiv},
       eprint = {0711.2267},
 primaryClass = {astro-ph},
       adsurl = {https://ui.adsabs.harvard.edu/abs/2008ApJ...676L..29H}
}

@ARTICLE{2025A&A...693L..10M,
       author = {{Mart{\'\i}nez-Sebasti{\'a}n}, C. and {Sim{\'o}n-D{\'\i}az}, S. and {Jin}, H. and {Keszthelyi}, Z. and {Holgado}, G. and {Langer}, N. and {Puls}, J.},
        title = "{The IACOB project: XIII. Helium enrichment in O-type stars as a tracer of past binary interaction}",
      journal = {A\&A},
         year = 2025,
        month = jan,
       volume = {693},
          eid = {L10},
        pages = {L10},
          doi = {10.1051/0004-6361/202452622},
archivePrefix = {arXiv},
       eprint = {2412.14107},
 primaryClass = {astro-ph.SR},
       adsurl = {https://ui.adsabs.harvard.edu/abs/2025A&A...693L..10M}
}

@ARTICLE{2021A&A...648A..65C,
       author = {{Castro}, N. and {Crowther}, P.~A. and {Evans}, C.~J. and {Vink}, J.~S. and {Puls}, J. and {Herrero}, A. and {Garcia}, M. and {Selman}, F.~J. and {Roth}, M.~M. and {Sim{\'o}n-D{\'\i}az}, S.},
        title = "{Mapping the core of the Tarantula Nebula with VLT-MUSE. II. The spectroscopic Hertzsprung-Russell diagram of OB stars in NGC 2070}",
      journal = {A\&A},
         year = 2021,
        month = apr,
       volume = {648},
          eid = {A65},
        pages = {A65},
          doi = {10.1051/0004-6361/202040008},
archivePrefix = {arXiv},
       eprint = {2102.03372},
 primaryClass = {astro-ph.SR},
       adsurl = {https://ui.adsabs.harvard.edu/abs/2021A&A...648A..65C}
}

@ARTICLE{2021ExA....51..887G,
       author = {{Garcia}, Miriam and {Evans}, Christopher J. and {Bestenlehner}, Joachim M. and {Bouret}, Jean Claude and {Castro}, Norberto and {Cervi{\~n}o}, Miguel and {Fullerton}, Alexander W. and {Gieles}, Mark and {Herrero}, Artemio and {de Koter}, Alexander and {Lennon}, Daniel J. and {van Loon}, Jacco Th. and {Martins}, Fabrice and {de Mink}, Selma E. and {Najarro}, Francisco and {Negueruela}, Ignacio and {Sana}, Hugues and {Sim{\'o}n-D{\'\i}az}, Sergio and {Sz{\'e}csi}, Dorottya and {Tramper}, Frank and {Vink}, Jorick S. and {Wofford}, Aida},
        title = "{Massive stars in extremely metal-poor galaxies: a window into the past}",
      journal = {Experimental Astronomy},
         year = 2021,
        month = jun,
       volume = {51},
       number = {3},
        pages = {887-911},
          doi = {10.1007/s10686-021-09785-x},
       adsurl = {https://ui.adsabs.harvard.edu/abs/2021ExA....51..887G}
}

\end{document}